\documentclass[iop,referee]{emulateapj-rtx4}

\usepackage{graphicx}  
\usepackage{dcolumn}   
\usepackage{bm}        
\usepackage{amsfonts,amsmath,amssymb,mathrsfs}
\usepackage{color}
\usepackage{hyperref}
\hypersetup{
  colorlinks=true,        
  linkcolor=blue,         
  citecolor=cyan,         
}

\usepackage{natbib,times}
\newcommand{\cf}{cf.~}
\newcommand{\ie}{i.e.,~}
\newcommand{\eg}{e.g.,~}

\shorttitle{Magnetically driven winds and X-ray afterglows of SGRBs}
\shortauthors{D. M. Siegel, R. Ciolfi, L. Rezzolla}

\begin{document}

\title{Magnetically driven winds from differentially rotating neutron stars\\
and X-ray afterglows of short gamma-ray bursts}

\author{Daniel M. Siegel\altaffilmark{1}, Riccardo Ciolfi\altaffilmark{1},
  and Luciano Rezzolla\altaffilmark{1,2}}

\altaffiltext{1}{Max Planck Institute for Gravitational Physics (Albert
  Einstein Institute), Am M\"uhlenberg 1, D-14476 Potsdam-Golm, Germany}

\altaffiltext{2}{Institut f\"ur Theoretische Physik, Max-von-Laue-Str. 1,
  D-60438 Frankfurt am Main, Germany}

\begin{abstract}
Besides being among the most promising sources of gravitational waves,
merging neutron star binaries also represent a leading scenario to
explain the phenomenology of short gamma-ray bursts (SGRBs). Recent
observations have revealed a large subclass of SGRBs with roughly
constant luminosity in their X-ray afterglows, lasting
$10\!-\!10^4\,\mathrm{s}$. These features are generally taken as evidence of
a long-lived central engine powered by the magnetic spin-down of a
uniformly rotating, magnetized object. We propose a different scenario in
which the central engine powering the X-ray emission is a differentially
rotating hypermassive neutron star (HMNS) that launches a quasi-isotropic
and baryon-loaded wind driven by the magnetic field, which is built-up through
differential rotation. Our model is supported by long-term,
three-dimensional, general-relativistic, and ideal magnetohydrodynamic
simulations, showing that this isotropic emission is a very robust
feature. For a given HMNS, the presence of a collimated component depends
sensitively on the initial magnetic field geometry, while the stationary
electromagnetic luminosity depends only on the magnetic energy initially
stored in the system. We show that our model is compatible with the
observed timescales and luminosities and express the latter in terms of a
simple scaling relation.
\end{abstract}
		
\keywords{gamma-ray burst: general -- magnetohydrodynamics (MHD) -- methods:
  numerical -- stars: magnetic field -- stars: neutron}

\section{Introduction}

Binary neutron star (BNS) mergers represent a leading scenario to
generate the physical conditions necessary for the launch of short gamma-ray bursts
(SGRBs; see, \eg \citealt{Paczynski86, Eichler89, Narayan92,
  Barthelmy2005, Fox2005,Gehrels2005,
  Rezzolla:2011,Berger2013,Tanvir2013}). Additionally, the inspirals and
mergers of BNSs are promising sources for the detection of gravitational
waves (GWs) with advanced interferometers such as LIGO and Virgo
\citep{Harry2010,Accadia2011} that have a predicted rate of
$40\,\mathrm{yr}^{-1}$ \citep{Abadie:2010}. Coincident electromagnetic
(EM) and GW observations are needed to confirm the association of SGRBs
with BNS mergers and to unravel the physical processes involved. Despite
observational evidence in some cases (\eg
\citealt{Burrows2006,Soderberg2006,Berger2013b}) and a requirement coming
from modeling merger rates \citep{Metzger2012}, it is not known whether
the prompt gamma-ray emission of an SGRB is always collimated in a
relativistic jet. If it was, a large fraction of SGRB events would be
missed, as their jets would be beamed away from us. Therefore,
understanding the potential EM signatures from BNS mergers and, in
particular, the non-collimated emission (\eg \citealt{Yu2013}), is
essential.

The \emph{Swift} satellite \citep{Gehrels2004} has recently revealed a large
subclass of SGRBs that show phases of roughly constant luminosity in
their X-ray afterglow lightcurves, referred to as ``extended emission''
and ``X-ray plateaus'' (\eg \citealt{Rowlinson2013, Gompertz2014}). The
associated X-ray emission can last from $10$ to $10^4\,\mathrm{s}$ and
the fluence can be comparable to or larger than that of the prompt emission.
These features are taken as a signature of a long-lived central engine as
proposed in \citet{Zhang2001}, where such emission is assumed to be
powered by the magnetic spin-down of a uniformly rotating object formed
in a BNS merger. In general, depending on the total mass and mass ratio
of the binary, the binary-merger product (BMP) can either be a
stable neutron star, a supramassive neutron star (SMNS, \ie a star with
mass above the maximum mass for nonrotating configurations $M_{_{\rm
    TOV}}$, but below the maximum mass for uniformly rotating
configurations $M_{\rm max}$, with $M_{\rm
  max}\simeq(1.15\!-\!1.20)\,M_{_{\rm TOV}}$,~\citealt{Lasota1996}), a
hypermassive neutron star (HMNS, \ie a star more massive than a SMNS), or
a black hole.\footnote{SMNSs or HMNSs will eventually collapse to a black
  hole, but on timescales that can be much larger than the dynamical
  one.}

In this Letter, we propose a different scenario to explain the X-ray
emission soon after the merger, in which the central engine is a
long-lived differentially rotating BMP and, in particular, an HMNS. Recent
observations of neutron stars with masses as large as $\simeq2\,M_\odot$
\citep{Demorest2010,Antoniadis2013} indicate a rather stiff equation of
state, while the mass distribution in BNSs is peaked around
$1.3\!-\!1.4\,M_\odot$ \citep{Belczynski2008}. This combined evidence
suggests that the BMP is almost certainly an SMNS or an HMNS.

On the basis of these considerations and using three-dimensional (3D)
general-relativistic ideal magnetohydrodynamic (MHD) simulations, we
investigate the EM emission of an initially axisymmetric HMNS endowed
with different initial magnetic fields spanning the range of reasonable
configurations. In all cases, we find a very luminous, quasi-stationary
EM emission from the HMNS, which is associated with a baryon-loaded outflow,
driven by magnetic winding in the stellar interior. The emission,
which is compatible with the observed X-ray afterglows, is
almost isotropic, though a collimated, mildly relativistic flow can be
produced if the magnetic field is mainly oriented along the spin axis.
The isotropic emission is present even in random-field, initial
configurations, making it a robust feature of a BNS merger and an
important EM counterpart to the GW signatures.

\section{Physical system and numerical setup}

As a typical HMNS, resulting from a BNS merger, we consider an
axisymmetric initial model constructed using the \texttt{RNS} code
\citep{Stergioulas95}, assuming a polytropic equation of state
$p=K\rho^{\Gamma}$, where $p$ is the pressure, $\rho$ the rest-mass
density, $\Gamma=2$, and $K=2.124\times 10^{5}\,{\rm cm}^5{\rm
  g}^{-1}{\rm s}^{-2}$.  The corresponding maximum gravitational mass for
a uniformly rotating (nonrotating) model is
$\approx2.27\,(1.97)\,M_{\odot}$. Our initial HMNS has a mass of
$M=2.43\, M_\odot$, an equatorial radius of $R_e=11.2\,\mathrm{km}$, and
it is differentially rotating according to a ``$j$-constant'' law
\citep{Komatsu89}, with the differential rotation parameter $A/R_e=1.112$ and
central period $P_c=0.47\,\mathrm{ms}$.

To study the influence of the magnetic field geometry on the EM emission,
we endow this model in hydrodynamic equilibrium with three different
initial magnetic field geometries, which are shown in the left panels of
Figures~\ref{fig:evolution_panels_1}--\ref{fig:evolution_panels_3}. The
first model (hereafter \texttt{dip-60}) employs a dipolar configuration
very similar to the one in \cite{Shibata2011b} and \cite{Kiuchi2012b},
specified by the azimuthal component of the vector potential
$A_\phi=A_{0,d}\varpi^2/(r^2+\varpi_{0,d}^2/2)^{3/2}$, where $\varpi$ is
the cylindrical radius, $r^2\equiv\varpi^2+z^2$, $A_{0,d}$ tunes the
overall field strength, and
$\varpi_{0,d}\simeq5.3\,R_e\simeq60\,\mathrm{km}$ indicates the radial
location of the magnetic neutral point on the equatorial plane. The
second model (henceforth \texttt{dip-6}) is endowed with the same field
geometry, but with $\varpi_{0,d}\simeq 0.53\,R_e\simeq6\,\mathrm{km}$. In
contrast to model \texttt{dip-60}, where the star is threaded by a
magnetic field that is artificially uniform on lengthscales $\gtrsim~R_e$
(\cf inset in Figure~\ref{fig:evolution_panels_1}), for model
\texttt{dip-6}, the neutral point and maximum magnetic field are located
in the interior of the star, corresponding to a more realistic
distribution of currents. Finally, the third model (henceforth
\texttt{rand}) represents a ``random'' magnetic field, which we compute
from a vector potential built as the linear superposition of
$\sim6\times10^4$ modes with random amplitudes and phases
\begin{eqnarray} 
\boldsymbol{A}_{ijk} &=& \frac{A_{0,r}\sqrt{\gamma}}{(r^2+\varpi_{0,r}^2)^{3/2}} 
\hskip -0.08cm 
\sum_{\ell m n=0}^{n_k} \mskip-5mu \!\!
\boldsymbol{a}_{\ell m n}\!\cos\left(\boldsymbol{x}_{ijk} \!\cdot\!
\boldsymbol{k}_{\ell m n} \!+\! 
2\pi \boldsymbol{b}_{\ell m n}\right)\nonumber \\ 
&&+\boldsymbol{c}_{\ell m n}\!\sin\left(\boldsymbol{x}_{ijk} \!\cdot\!
\boldsymbol{k}_{\ell m n} \!+\! 2\pi 
\boldsymbol{d}_{\ell m n}\right)\,. 
\end{eqnarray} 
Here, ${i,j,k}$ label the points on the computational grid,
$\boldsymbol{x}_{ijk}$ denotes the associated position vectors, and
$\ell,\,m,\,n$ refer to a grid in wave vector $\boldsymbol{k}$ space,
defined by
$\{\boldsymbol{k}_{\ell m n}\}\equiv[0,2\pi/\lambda_\mathrm{min},2\pi/(\lambda_\mathrm{min}+\Delta\lambda),\ldots,2\pi/(\lambda_\mathrm{min}+(n_k-1)\Delta\lambda)]^3$,
where $n_k=30,~\lambda_\mathrm{min}\approx3\,\mathrm{km}$ is the smallest
wavelength employed, and
$\Delta\lambda\approx2.2\,\mathrm{km}$. Furthermore,
$\boldsymbol{a}_{\ell m n},\ldots,\boldsymbol{d}_{\ell m n}$ are random
numbers between 0 and 1. The constant $A_{0,r}$ sets the maximum field
strength, the factor $\sqrt{\gamma}$, with $\gamma$ as the determinant of
the spatial metric, helps to concentrate stronger magnetic fields in
regions of larger space--time curvature, while the denominator scales the
magnetic field as $\sim\!r^{-3}$ for $r>\varpi_{0,r}\approx2\,R_e$. We
built the random-field model to resemble the actual magnetic field
configuration resulting from a BNS merger (\cf
\citealt{Rezzolla:2011}). In all the above cases, $A_{0,d}$ and $A_{0,r}$
are adjusted to yield maximum initial field strengths of $B_0=2\times
10^{14}\,\mathrm{G}$, with the magnetic-to-fluid pressure ratio being
$\ll\!10^{-5}$ inside the star (cf. Figure~\ref{fig:pb_pg}). The resulting
initial magnetic energy $E_{_\mathrm{M}}$ differs with the magnetic field
geometry, being
$E_{_\mathrm{M}}\simeq(530,\,1.5,\,5.1)\times10^{44}\,\mathrm{erg}$ for
models \texttt{dip-60}, \texttt{dip-6} and \texttt{rand}, respectively.

The time evolution of these initial data is performed using the publicly
available Einstein Toolkit with the \texttt{McLachlan}
spacetime-evolution code \citep{loeffler_2011_et}, combined with the
fully general-relativistic ideal-MHD code \texttt{WhiskyMHD}
\citep{Giacomazzo:2007ti}. The ideal-fluid equation of state
$p=\rho\epsilon(\Gamma-1)$ is used for the evolution, where $\epsilon$ is
the specific internal energy and $\Gamma=2$. The star is initially
surrounded by a low-density atmosphere with
$\rho_\mathrm{atm}\simeq6\times10^{-9}\rho_\mathrm{c}$, where
$\rho_\mathrm{c}\simeq1.0\times10^{15}\,\mathrm{g\,cm^{-3}}$ is the
central rest-mass density. We carry out the MHD computations within the
modified Lorenz gauge approach \citep{Farris2012,Giacomazzo2013}, with
damping parameter $\xi=2/M$. The computational grid consists of a
hierarchy of seven nested boxes, extending to $\approx
105\,R_e\approx1180\,\mathrm{km}$ in the $x$- and $y$-directions and to
$\approx72\,R_e\approx817\,\mathrm{km}$ in the $z$-direction. The finest
refinement level corresponds to a spatial domain of
$[0,18.4]\times[0,18.4]\times[0,12.8]\,\mathrm{km}$ and covers the HMNS
at all times. The highest spatial resolution is $h\simeq 140\,{\rm m}$,
but lower-resolution simulations have been performed to check
convergence. To make these long-term simulations computationally
affordable in 3D and at high resolution, a $\pi/2$ rotation symmetry
around the $z$-axis and a reflection symmetry across the $z=0$ plane have
been employed.

\section{Results}

\begin{figure*} 
  \centering   
  \includegraphics[angle=0,width=0.98\textwidth]{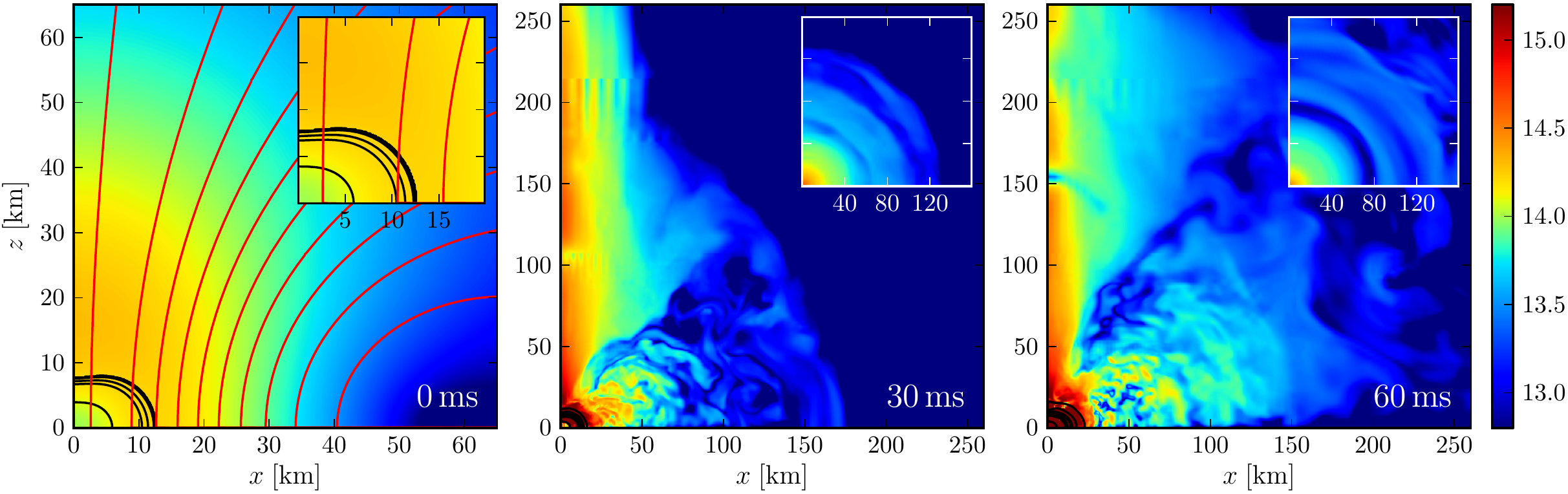} 
  \caption{Snapshots of the magnetic field strength (color-coded in
  logarithmic scale and Gauss) and rest-mass density contours in
  the $(x,z)$ plane at representative times for model
  \texttt{dip-60}. Magnetic field lines are drawn in red in the
  left panel. The leftmost inset shows a magnification of
  the HMNS, the other ones show a horizontal cut at $z=120\,\mathrm{km}$.}
  \label{fig:evolution_panels_1}
\end{figure*}

\begin{figure*} 
  \centering
  \includegraphics[angle=0,width=0.98\textwidth]{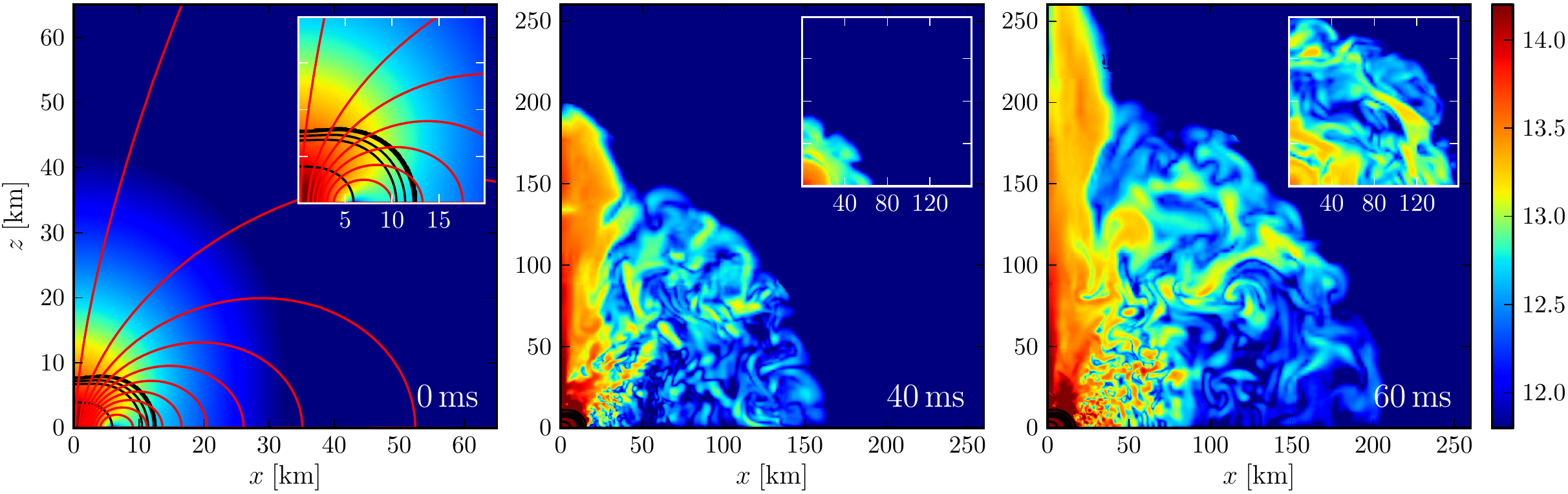} 
  \caption{Same as Figure~\ref{fig:evolution_panels_1}, but for model \texttt{dip-6}.}
  \label{fig:evolution_panels_2} 
\end{figure*}

\begin{figure*} 
  \centering
  \includegraphics[angle=0,width=0.98\textwidth]{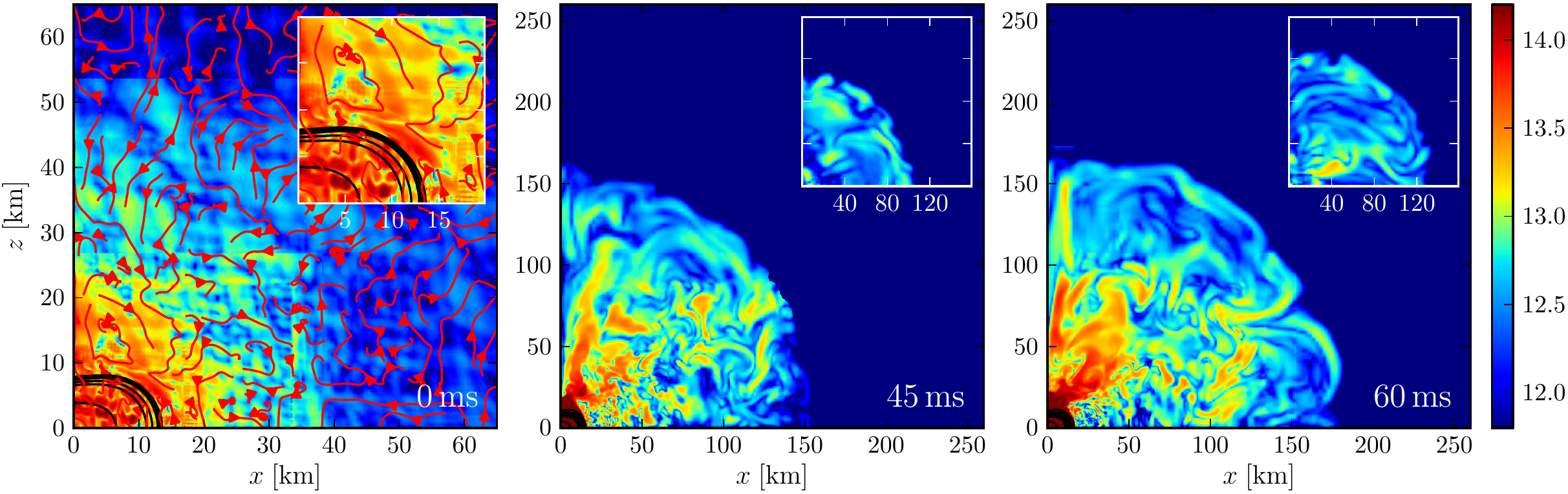} 
  \caption{Same as Figure~\ref{fig:evolution_panels_1}, but for model \texttt{rand}.}
  \label{fig:evolution_panels_3} 
\end{figure*}

\begin{figure*} 
  \centering   
  \includegraphics[angle=0,width=0.98\textwidth]{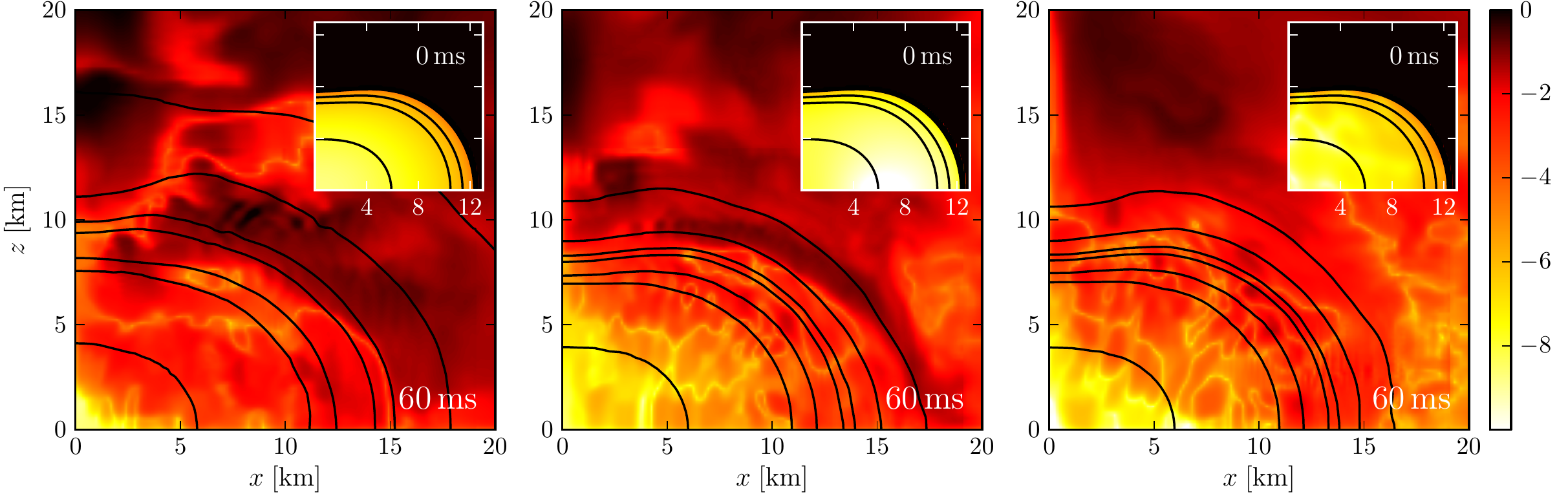} 
  \caption{Logarithmic magnetic-to-fluid pressure ratio and rest-mass
    density contours for model \texttt{dip-60}
    (left), \texttt{dip-6} (middle), and \texttt{rand} (right).}
  \label{fig:pb_pg}
\end{figure*}

Figures~\ref{fig:evolution_panels_1}--\ref{fig:evolution_panels_3} show
snapshots of the norm of the magnetic field and rest-mass density
contours in the $(x,z)$ plane for the three models discussed above at
representative times during the evolution. Due to the differential
rotation in the HMNS, strong toroidal fields are generated via magnetic
winding (linearly growing at the beginning). Within a few rotational
periods, the build-up of magnetic pressure (mostly associated with
steep toroidal-field radial gradients) is
sufficient to overcome the gravitational binding in the vicinity of the
stellar surface, powering an outflow of highly magnetized low-density
matter with mildly relativistic velocities (cf. also
Figure~\ref{fig:pb_pg}).


For model \texttt{dip-60}, since the initial magnetic field is oriented
along the rotation axis on lengthscales $\gtrsim~R_e$ (\cf
Figure~\ref{fig:evolution_panels_1}, left panel), the outflow is
efficiently channeled along this axis and thus significantly collimated,
though this is more the result of the ordered initial magnetic field
configuration than of genuine self-confining processes. During the
evolution, the non-collimated part of the outflow at lower latitudes
tends to push the collimated part toward the axis, effectively shrinking
the inner opening angle of this jet-like structure (\cf
Figure~\ref{fig:evolution_panels_1}). This triggers irregularities in the
collimated outflow that manifest themselves as small bubbles of
material with low density and magnetization being released along the
vertical axis, possibly pointing to a kink
instability~\citep{Kiuchi2012b}.

The initial magnetic field geometry of \texttt{dip-6} has a much smaller
curvature scale and is less effective at funneling the outflow; the
resulting collimation is thus much less pronounced than for
\texttt{dip-60}. Furthermore, while model \texttt{dip-60} shows a clear
distinction between the collimated and the non-collimated part of the
outflow, this is no longer the case for model \texttt{dip-6} (\cf
Figures~\ref{fig:evolution_panels_1} and \ref{fig:evolution_panels_2}).
Finally, the model with the most realistic magnetic field configuration,
\ie model \texttt{rand}, is characterized by an almost isotropic outflow
(\cf Figure~\ref{fig:evolution_panels_3}), though we cannot exclude
the fact that
some form of collimation could emerge at later times.  This represents an
important result of our simulations: the amount of collimation in the
magnetic-driven wind from the BMP will depend sensitively on the magnetic
field geometry and could be absent if the field is randomly distributed.

In all of the configurations considered, the magnetized baryon-loaded
outflow has rest-mass densities $\sim10^{8}\!-\!10^{9}\,{\rm g\,cm}^{-3}$ and
is ejected from the star with velocities $v/c\lesssim0.1$, in the
isotropic part, and $v/c\lesssim0.3$, in the collimated part.

\begin{figure*} 
\centering 
\includegraphics[angle=0,width=0.48\textwidth]{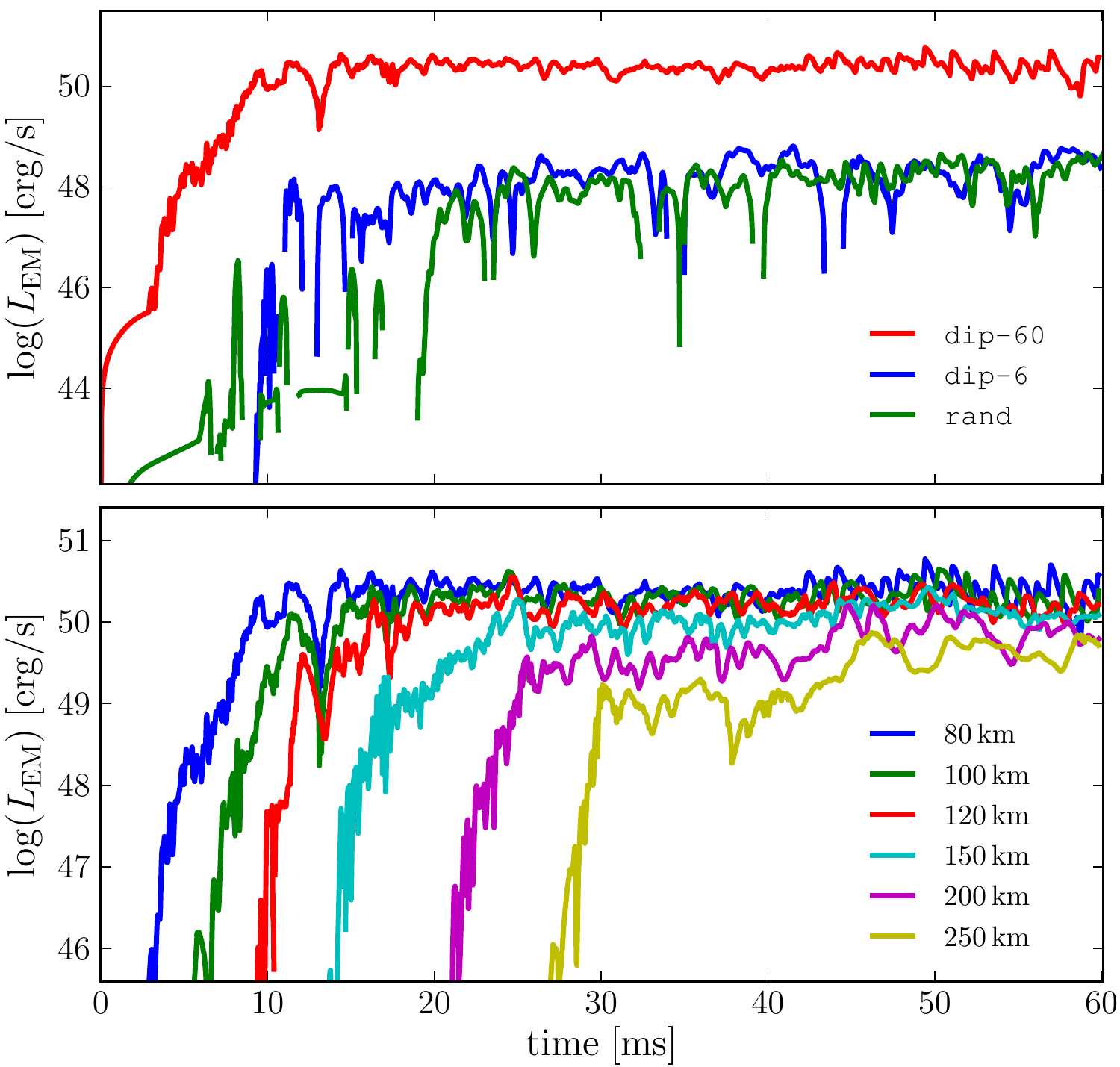} 
\hskip 0.5cm 
\includegraphics[angle=0,width=0.48\textwidth]{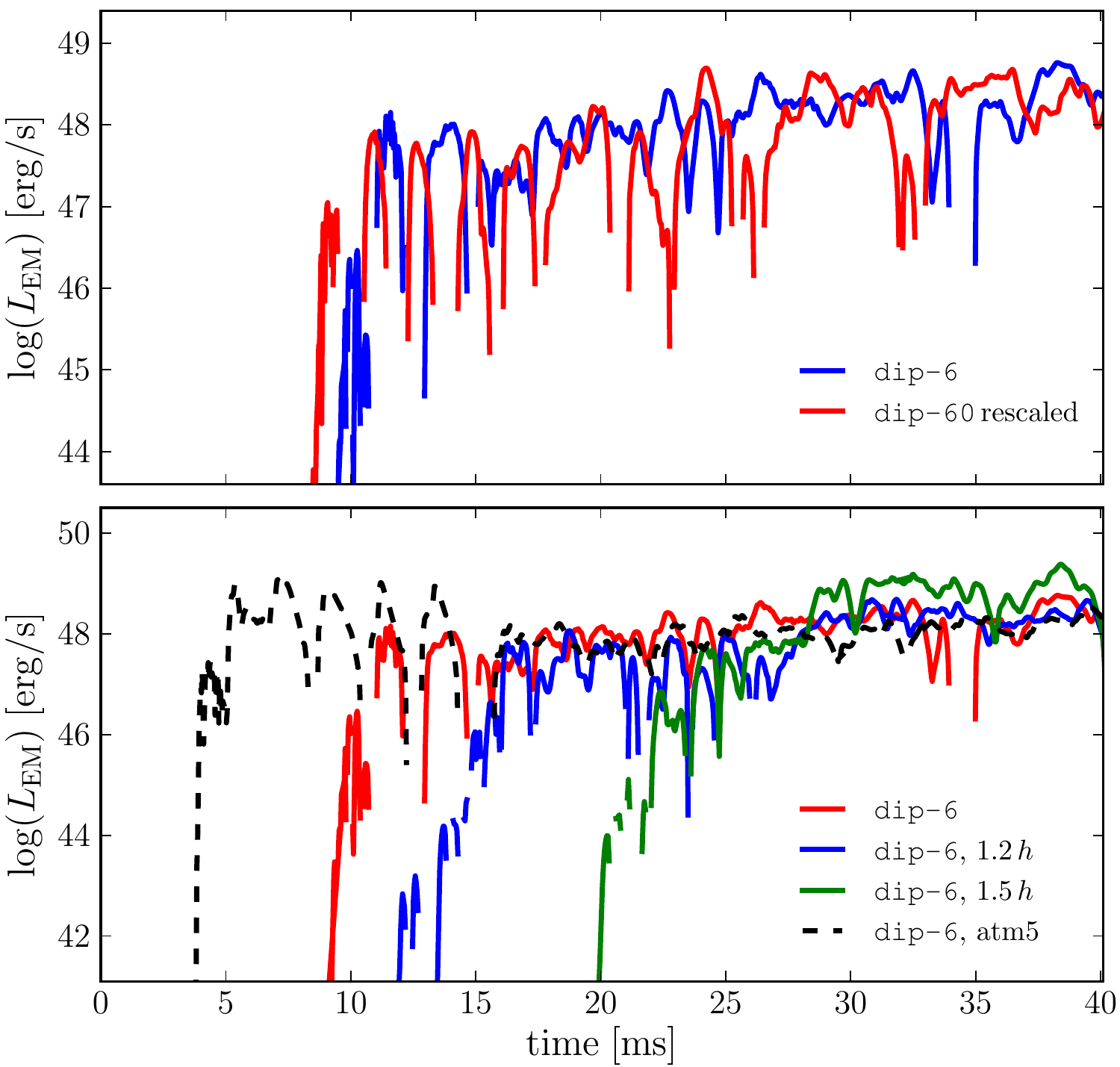} 
\caption{Top left: EM luminosities for the different simulations. Top
  right: luminosity comparison between model \texttt{dip-6} and model
  \texttt{dip-60} with rescaled initial magnetic field to match the
  initial magnetic energy of model \texttt{dip-6}. Bottom left: luminosities
  extracted at different radii for model \texttt{dip-60}. Bottom right:
  impact of resolution and a factor of 10 lower atmosphere threshold on
  simulations of model \texttt{dip-6}.}
\label{fig:luminosities} 
\end{figure*}

Defining the isotropic luminosity as
\begin{equation} 
  L_{_\mathrm{EM}} \equiv -\oint_{r=R_\mathrm{d}} \mskip-20mu d\Omega\,\sqrt{-g}\, 
  (T^{^{\mathrm{EM}}})^{r}_{\phantom{r}t}\,,
\label{eq:lum}\end{equation} 
where $d\Omega$ is the solid-angle element, $g$ is the determinant of the
spacetime metric, and $T^{^{\mathrm{EM}}}_{\mu\nu}$ is the EM part of the
energy-momentum tensor, all of the different initial magnetic field
geometries yield high, stationary EM luminosities associated with the
outflow of
$L_{_\mathrm{EM}}\sim10^{48}\!-\!10^{50}\,\mathrm{erg}\,\mathrm{s}^{-1}$ (\cf
Figure~\ref{fig:luminosities}, upper left panel). Model \texttt{dip-60} has
a luminosity that is two orders of magnitude larger than those of models
\texttt{dip-6} and \texttt{rand}, which have surprisingly similar
luminosities. These differences can be understood in terms of the
different magnetic energies associated with the three models (see below).

We have also verified that when computed at sufficiently large distances,
the luminosity depends only weakly on the radius $R_\mathrm{d}$, chosen to
compute the integral in Equation~\eqref{eq:lum} (see
Figure~\ref{fig:luminosities}, lower left panel). The remaining
discrepancies among different $R_\mathrm{d}$ choices are due to the bulk
of the non-collimated part of the outflow having not yet crossed the
outer detection spheres at $t=60~\mathrm{ms}$. The effective bulk speed
of the outflow at larger distances (as inferred from the time delay in
the onset of $L_{_\mathrm{EM}}$ at different $R_\mathrm{d}$) is smaller
than the ejection speed of $\lesssim0.1\,c$, due to a nonzero atmosphere
rest-mass density floor. This effect is illustrated in the lower right
panel of Figure~\ref{fig:luminosities}, where a lower floor yields a higher
bulk speed (\cf black and red lines). Note, however, that the stationary
luminosity is not affected by the atmosphere level even when the latter
is changed by one order of magnitude. Furthermore, the lower right panel
of Figure~\ref{fig:luminosities} reports the luminosity for three different
resolutions, showing that the outflow speed is affected but the
stationary levels are in reasonably good agreement.

Another important conclusion to be drawn from our simulations is that,
keeping the background fluid model fixed, the EM luminosity
essentially depends on the initial magnetic energy of the system. In this
way, the difference of two orders of magnitude between the luminosities of
model \texttt{dip-60} and those of models \texttt{dip-6} and \texttt{rand} (\cf
upper left panel of Figure~\ref{fig:luminosities}) is simply due to the
corresponding difference in $E_{_\mathrm{M}}$. This is demonstrated in
the upper right panel of Figure~\ref{fig:luminosities}, where we show the
luminosity of a model \texttt{dip-60} with an initial magnetic field
strength that is rescaled to match the initial magnetic energy of
model \texttt{dip-6}. Once the initial magnetic energies coincide, the
resulting stationary luminosities are the same and therefore
independent of the magnetic field geometry and degree of collimation.

Our results lead to a simple expression for the EM luminosity from the
BMP,
\begin{equation} 
  L_{_\mathrm{EM}} \! \simeq \!
  10^{48} \chi \left(\frac{B_0}{10^{14}\,\mathrm{G}}\right)^{\!\!2} \!\!
\left(\frac{R_e}{10^{6}\,\mathrm{cm}}\right)^{\!\!3} \!\!
\left(\frac{P}{10^{-4}\,\mathrm{s}}\right)^{\!\!-1} \!\!\!\!
\mathrm{erg}\,\mathrm{s}^{-1} \!, 
\label{eq:scaling}
\end{equation} 
where $P$ denotes the (central) spin period and $B_0$ denotes the initial maximum
magnetic field. Our simulations reveal the need for a fudge factor, $\chi$,
that is $\sim\!1$ for the more realistic magnetic field configurations
\texttt{rand} and \texttt{dip-6}, and $\sim\!100$ for \texttt{dip-60},
showing that initial models with the same $B_0$ but different field
geometries lead to luminosities that differ by orders of magnitude (\cf
upper left panel of Figure~\ref{fig:luminosities}). The role of the
equatorial radius, $R_e$, is to define the volume in which the conversion
of rotational energy into EM energy occurs, while the period, $P$, sets the
timescale of such conversion.

Equation~\eqref{eq:scaling} still holds if expressed in terms of
quantities referring to the stage when the outflow has become roughly
stationary, by substituting $\chi\,(B_0/10^{14}\,\mathrm{G})^2$ with
$(\bar{B}/10^{15}\,\mathrm{G})^2$, where $\bar{B}$ is the magnetic field
strength reached in the outer layers of the star ($R_e$ and $P$ are
essentially unchanged from $t=0$). The fudge factor is no longer
necessary, as this new expression does not depend on the magnetic field
geometry.

Two remarks should be made. First, scalings similar to
Equation~\eqref{eq:scaling} have already been reported in the literature (\eg
\citealt{Meier1999,Shibata2011b}), where, however, no discussion was made
on the dependence of the luminosity on the initial magnetic field
geometry and thus on the importance of the factor $\chi$. Ignoring this
dependence can easily lead to over/underestimates of the luminosity by
orders of magnitude. Second, the scaling with radius and period in
Equation~\eqref{eq:scaling} are very different from those resulting from the
magnetic dipole spin-down emission considered in the model of
\citet{Zhang2001}, where the luminosity is given by
$L_{_\mathrm{EM}}\propto B^2\,R_e^6\,P^{-4}$. These differences in
scaling may be used to discriminate between the scenario proposed here
and that by \citet{Zhang2001}.

\section{Magnetic-driven winds and X-ray emission}

Defining the conversion efficiency of EM luminosity into observed X-ray
luminosity as
$\eta\equiv\,L_{_\mathrm{EM}}^{\mathrm{obs}}/L_{_\mathrm{EM}}$, we can
use the results of our simulations for characteristic radii and periods
($R_{e}\sim10^{6}\,{\rm cm}$ and $P\sim10^{-4}\,{\rm s}$) to deduce that
magnetic field strengths in the range
$\bar{B}\sim\eta^{-1/2}\times10^{14}\!-\!10^{16}\,\mathrm{G}$ are needed to
produce luminosities
$L_{_\mathrm{EM}}^{\mathrm{obs}}\sim10^{46}\!-\!10^{51}\,\mathrm{erg}\,\mathrm{s}^{-1}$,
which characterize the extended emission and X-ray plateaus of SGRBs (\eg
\citealt{Rowlinson2013,Gompertz2014}).  Assuming an efficiency of, e.g.,
$\eta\sim0.01\!-\!0.1$ yields magnetic field strengths
$\bar{B}\sim10^{14}\!-\!10^{17}\,\mathrm{G}$. These strengths are not reached
in the progenitor neutron stars, but they can be built from much weaker
initial magnetic fields in a number of ways: the compression of stellar
cores \citep{Giacomazzo:2010}, the Kelvin--Helmholtz instability
developing during the merger (\citealt{Price06, Baiotti08, Anderson2008,
  Giacomazzo:2010, Zrake2013b}), magnetic winding and the
magnetorotational instability (MRI; \citealt{Siegel2013}). In particular,
the MRI might enhance the EM emission, though its saturation level is
unknown and its role in determining luminosity levels remains
unclear. Unfortunately, resolving the MRI in long-term global simulations
is out of reach. In our simulations only magnetic winding is at play,
yet magnetic fields are amplified by at least one order of magnitude.

The timescale for the persistence of differential rotation and the
survival time of the BMP is still very
much uncertain \citep{Lasky2014} and it is hard to estimate the timescale
$\tau_{_{\mathrm{EM}}}$ over which these luminosities can be
sustained. Simulations show that the timescale for the extraction of
angular momentum via GWs from a bar-deformed BMP is $\lesssim1\,{\rm s}$
(\citealt{Baiotti08}; \citealt{Bernuzzi2013}; but see also \citealt{Fan2013}). Hence, if the BMP survives for more
than $\sim1\,{\rm s}$, it must have been driven toward an almost
axisymmetric equilibrium by GW emission. A back of the envelope estimate
of magnetic braking leads to a timescale of $\sim1\!-\!10$~s for the magnetic
fields considered here \citep{Shapiro00}. This is confirmed by the
evolution of the angular velocity in our simulations, which shows that
$\Omega/\dot{\Omega}\sim10$~s in the stellar interior. Both of these
timescales should be meant as lower limits and differential rotation will
be removed on timescales that could be 10 times larger. However, even
100 s are not sufficient to explain X-ray luminosities lasting
$10^3\!-\!10^4$ s. A viable scenario is one in which the BMP is an HMNS at
birth, but evolves into an SMNS while differential rotation is removed via
magnetic braking and rest mass is lost via the wind (our simulations show
mass-loss rates of $\sim10^{-3}\!-\!10^{-2}\,M_{\odot}$\,s$^{-1}$). In this
case, the BMP can survive on much longer timescales as it needs less
angular momentum to prevent gravitational collapse. Once differential
rotation has been removed (or is very small), the spin-down via dipolar
emission through the global magnetic field produced over these timescales
would power the EM emission \citep{Zhang2001,Gao2006}. Note that a similar evolution also applies
if the differentially rotating BMP is an SMNS at birth.

Considering a reference luminosity of
$L_{_\mathrm{EM}}\sim\,10^{48}\,\mathrm{erg}\,\mathrm{s}^{-1}$, the
timescale needed to exhaust the reservoir of rotational energy
$T\sim5\times10^{52}\,\mathrm{erg}$ of our BMP is readily given by
$\tau_{_{\mathrm{EM}}}\lesssim\,T/L_{_\mathrm{EM}}\sim5\times10^4\,\mathrm{s}$. Through
fitting the data reported in Table 3 of \citet{Rowlinson2013} we find a
power-law correlation between the observed plateau luminosities and
durations:
$L_{_\mathrm{EM}}^{\mathrm{obs}}\,[\mathrm{erg}\,\mathrm{s}^{-1}]\sim10^{52}\,(\tau_{_{\mathrm{EM}}}\,[\mathrm{s}])^{-a}$,
where $a=1.36\pm0.11$. Given a luminosity of
$L_{_\mathrm{EM}}=L_{_\mathrm{EM}}^{\mathrm{obs}}/\eta\sim10^{48}\,\mathrm{erg}\,\mathrm{s}^{-1}$,
the fit gives a duration of the plateau emission
$\tau_{_{\mathrm{EM}}}\sim(10^4/\eta)^{1/a}\,\mathrm{s}\sim5\times10^3\,{\rm
  s}$ for $\eta\sim0.1$. It is reassuring that even these crude estimates
provide emission timescales that are well below the upper limit of
$T/L_{_\mathrm{EM}}\sim5\times10^4\,\mathrm{s}$, thus showing the
compatibility with the observations.

We also note that the estimates made above are based on the assumption
that the observed emission is essentially isotropic. If there is
collimation within a solid angle $\Omega_\mathrm{coll}$, the duration
$\tau_{_{\mathrm{EM}}}$ estimated from the fit would be reduced by a
factor of $\sim(\Omega_\mathrm{coll}/4\pi)^{1/a}$.

\section{Conclusions}

Using ideal MHD simulations we have investigated the EM emission of an
initially axisymmetric HMNS endowed with different initial magnetic field
configurations, spanning the range of geometries expected from BNS
mergers. Despite the different initial configurations, we have found that,
in all cases, differential rotation in the HMNS generates a strong
toroidal magnetic field and a consequent baryon-loaded outflow with bulk
velocities $\lesssim0.1\,c$. This emission is almost isotropic, though
an additional collimated, mildly relativistic flow is produced if the
initial magnetic field has a dominant dipole component along the spin
axis.

Since the emission we observe emerges as a robust feature of a BNS merger
and given the consistency of the luminosity levels and duration with the
observations, we conclude that the proposed physical mechanism represents
a viable explanation for the X-ray afterglows of SGRBs.

\bigskip
We thank F. Pannarale and F. Galeazzi for discussions, W. Kastaun for
help with the visualizations, and the referee for valuable comments. R.C. is
supported by the Humboldt Foundation. Additional support comes from the
DFG Grant SFB/Transregio~7, from ``NewCompStar'', COST Action MP1304, and
from HIC for FAIR. The simulations have been performed on SuperMUC at LRZ
Garching and on Datura at the AEI.


\end{document}